\UseRawInputEncoding
\documentclass[]{spie}  

\usepackage{amsmath,amsfonts,amssymb}
\usepackage{subcaption}
\usepackage{graphicx}
\usepackage{cite} 
\usepackage{times}
\usepackage{epsfig}
\usepackage{amsmath}
\usepackage{nccmath}
\usepackage{amssymb}
\usepackage{mwe}
\usepackage{acro}
\usepackage{amssymb}
\usepackage{xcolor,colortbl}
\usepackage{tabularx}
\usepackage{relsize}
\usepackage{pifont}
\usepackage{booktabs} 
\usepackage{multirow}
\usepackage{multicol}
\usepackage{adjustbox}
\usepackage{float}
\usepackage{graphicx}
\usepackage{makecell}
\usepackage{tabu}
\usepackage[colorlinks=true, allcolors=blue]{hyperref}
\usepackage[capitalize]{cleveref}

\title{Leverage Weakly Annotation to Pixel-wise Annotation via Zero-shot Segment Anything Model for Molecular-empowered Learning}

\author[*a]{Xueyuan Li}
\author[*b]{Ruining Deng}
\author[c]{Yucheng Tang}
\author[d]{Shunxing Bao}
\author[e]{Haichun Yang}
\author[a,b,d,e]{Yuankai Huo}

\affil[a]{Data Science Institute, Vanderbilt University, Nashville, TN, USA}
\affil[b]{Department of Computer Science, Vanderbilt University, Nashville, TN, USA}
\affil[c]{NVIDIA Corporation, Redmond, WA, USA}
\affil[d]{Department of Electrical and Computer Engineering, Vanderbilt University Medical Center, Nashville, TN, USA}
\affil[e]{Department of Pathology, Microbiology and Immunology, Vanderbilt University Medical Center, Nashville, TN, 37232, USA}

\authorinfo{*Xueyuan Li and Ruining Deng contributed equally to this paper\\ Corresponding author: Yuankai Huo: E-mail: yuankai.huo@vanderbilt.edu}

\begin{document} 
\maketitle

\begin{abstract}
Precise identification of multiple cell classes in high-resolution Giga-pixel whole slide imaging (WSI) is critical for various clinical scenarios. Building an AI model for this purpose typically requires pixel-level annotations, which are often unscalable and must be done by skilled domain experts (e.g., pathologists). However, these annotations can be prone to errors, especially when distinguishing between intricate cell types (e.g., podocytes and mesangial cells) using only visual inspection. Interestingly, a recent study showed that lay annotators, when using extra immunofluorescence (IF) images for reference (referred to as molecular-empowered learning), can sometimes outperform domain experts in labeling. Despite this, the resource-intensive task of manual delineation remains a necessity during the annotation process. In this paper, we explore the potential of bypassing pixel-level delineation by employing the recent segment anything model (SAM) on weak box annotation in a zero-shot learning approach. Specifically, we harness SAM's ability to produce pixel-level annotations from box annotations and utilize these SAM-generated labels to train a segmentation model. Our findings show that the proposed SAM-assisted molecular-empowered learning (SAM-L) can diminish the labeling efforts for lay annotators by only requiring weak box annotations. This is achieved without compromising annotation accuracy or the performance of the deep learning-based segmentation.
This research represents a significant advancement in democratizing the annotation process for training pathological image segmentation, relying solely on non-expert annotators.

\end{abstract}



  


\section{INTRODUCTION}
\label{sec:intro}  
Accurate identification of multiple cell classes within high-resolution Giga-pixel whole slide images (WSI) is paramount across a broad spectrum of clinical scenarios, from disease diagnosis to treatment planning. The construction of an AI model for this task typically necessitates labor-intensive pixel-level annotation~\cite{wu2023diffumask,bonechi2020weak}, undertaken by domain specialists such as pathologists. However, this mode of annotation can be susceptible to errors, especially when discerning between intricate cell types (e.g., podocytes and mesangial cells) based solely on visual assessment.

Intriguingly, recent research has shown that non-specialist or lay annotators outperform domain experts in labeling accuracy when they refer to supplemental, annotation-free molecular information~\cite{deng2023democratizing}. This superior performance stems from the inclusion of additional immunofluorescence (IF) images, which harness molecular insights to augment learning~\cite{day2013fluorescently,moore2017effects}. Yet, despite this advancement, the resource-heavy manual delineation inherent to the annotation process remains indispensable, keeping the search for innovative methods to simplify or automate this step alive.

\begin{figure*}[t]
\begin{center}
\includegraphics[width=1\linewidth]{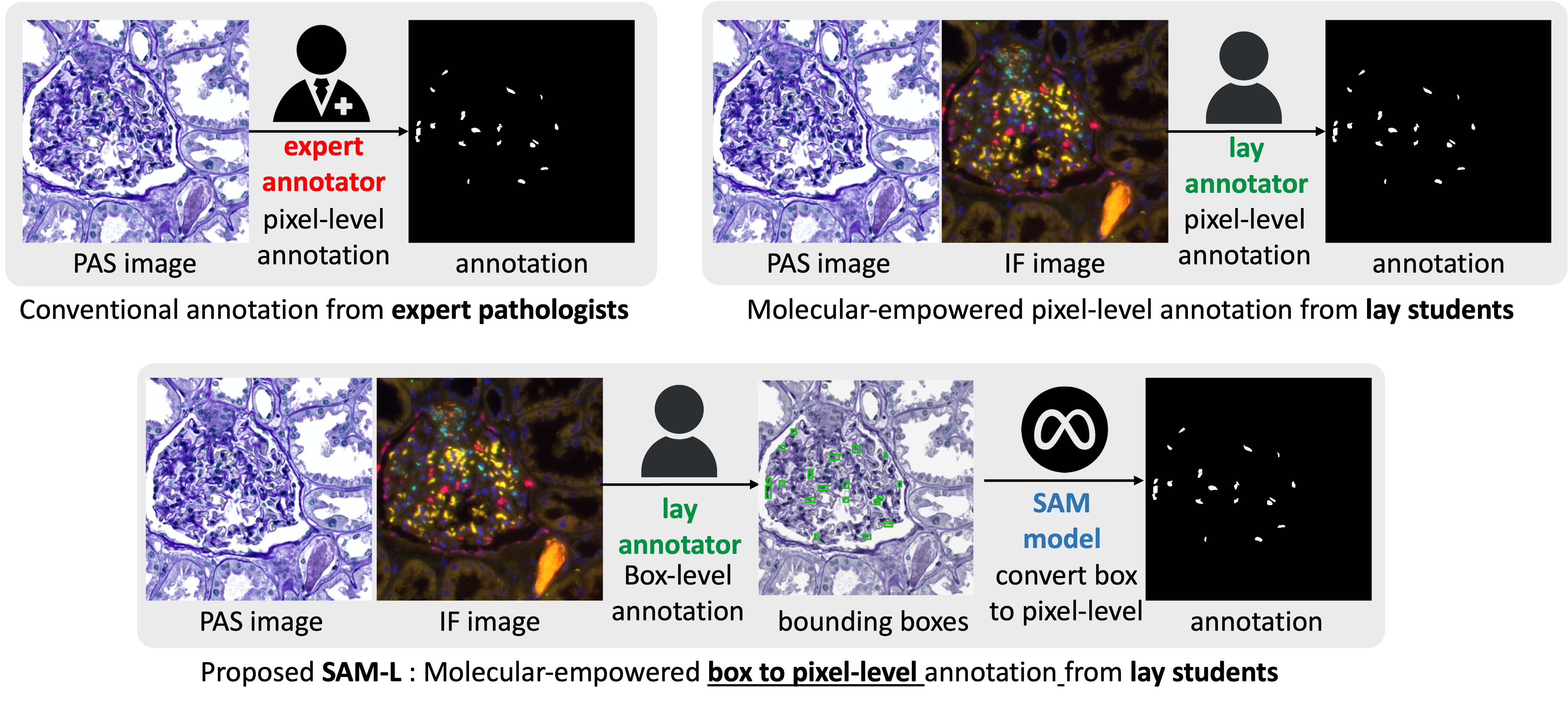}
\end{center}
\caption{\textbf{Major Idea of This Work.} The top-left panel illustrates the conventional annotation process employing only PAS images for pathological segmentation. In contrast, the top-right panel shows molecular-informed annotation utilizing both PAS and IF images, yielding superior annotation quality by lay annotators compared to the top-left process. The bottom panel demonstrates the proposed SAM-L annotation method, which leverages box annotations to accomplish pixel-level segmentation results, using these boxes as prompts for the zero-shot SAM segmentation. This holistic approach enhances annotation quality and paves the way for more precise and resilient segmentation models.}
\label{fig:idea}
\end{figure*}

This paper introduces the segment anything model (SAM)\cite{kirillov2023segment}, a tool that converts weak annotations (e.g., bounding boxes) into pixel-level annotations, utilizing a zero-shot learning strategy\cite{yu2019zero} for segmentation. The central concept of our study is presented in Fig.\ref{fig:idea}. Our SAM-assisted, molecular-empowered learning (SAM-L) method aims to curtail the labeling efforts required of lay annotators by soliciting only weak box annotations, yet it maintains the robust segmentation performance typical of deep learning methods. Notably, our training dataset comprises diverse Periodic Acid-Schiff (PAS)~\cite{aterman1963periodic,fu2017periodic} images and IF images of glomerulus slides, with only the PAS images being necessary during the testing/inference phase. Each training image received annotations from both experts and lay annotators. By integrating corrective learning into our strategy, we utilize diverse annotations from experts, students, and SAM's predictive annotation metrics to adjust and fine-tune initial segmentations. This pioneering blend promises to elevate cell segmentation and democratize pathology AI, signifying a harmonious integration of human expertise and technological advancement.

\begin{figure*}[t]
\begin{center}
\includegraphics[width=1\linewidth]{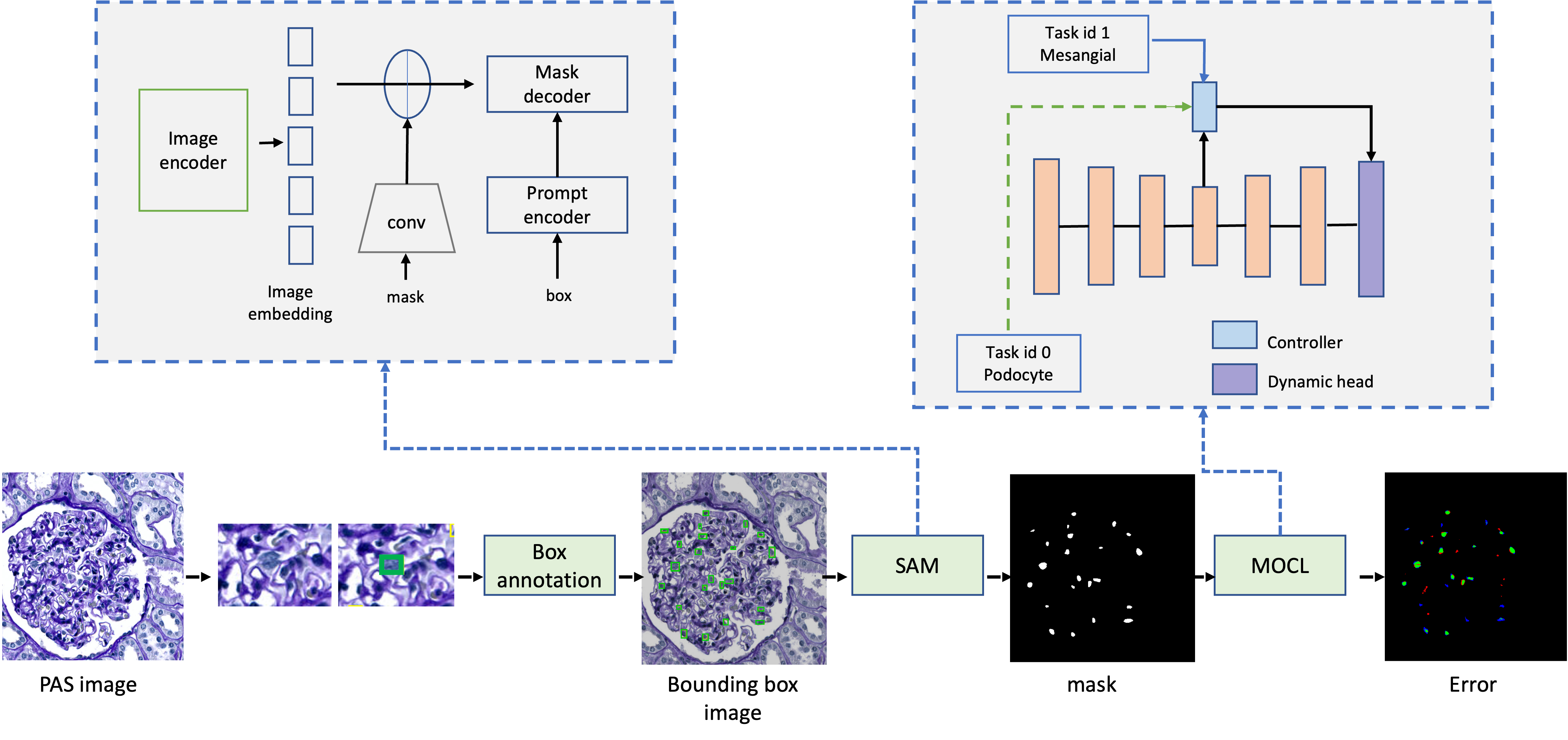}
\end{center}
\caption{\textbf{Framework of the SAM-L Scheme.} The SAM-assisted molecular-empowered learning pipeline comprises: (1) Box Annotation, (2) Segment Anything Model (SAM), and (3) Molecular-Oriented Corrective Learning (MOCL).}
\label{fig:framework}
\end{figure*}

\section{Method}
\subsection{Weak to Strong Annotation via SAM}
The SAM model~\cite{cui2023all,kirillov2023segment,deng2023segment} is a sophisticated deep learning model adept at generating high-quality object masks from diverse input prompts, such as points or boxes. It can produce masks for every object within an image. SAM was trained on a comprehensive dataset containing 11 million images and a staggering 1.1 billion masks. This rigorous training endowed SAM with exceptional zero-shot capabilities, making it extremely versatile for a myriad of segmentation challenges. Given its remarkable proficiency and thorough training, SAM stands as a notable milestone in deep learning-based image segmentation.

\begin{figure*}[t]
\begin{center}
\includegraphics[width=1\linewidth]{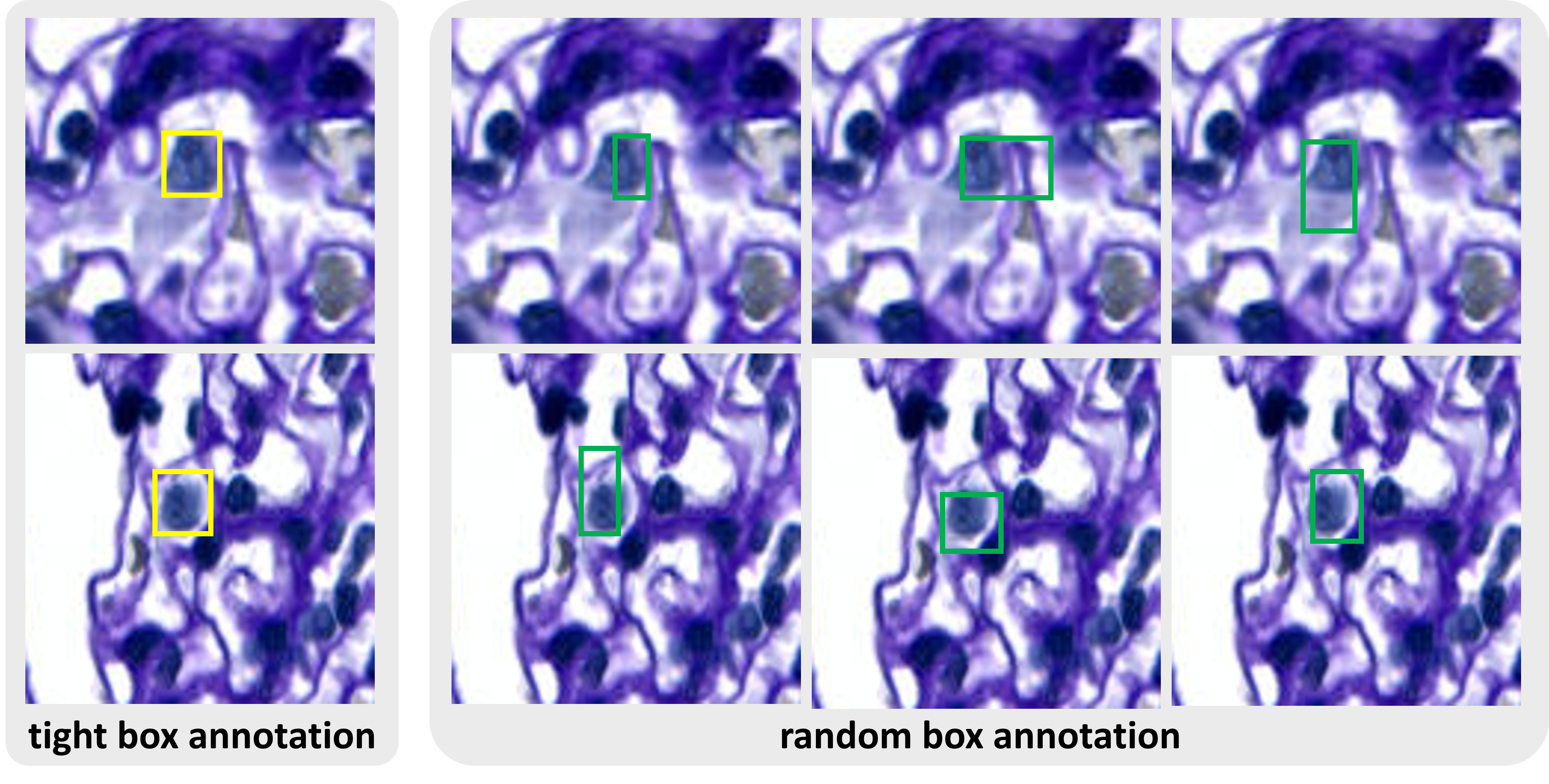}
\end{center}
\caption{\textbf{Tight Box vs. Random Box.} This figure illustrates two kinds of simulated boxes employed to produce pixel-level SAM annotations from weak human box annotations. Tight boxes are crafted as minimal-size boxes encompassing the original pixel-level annotation, whereas random boxes simulate the more realistic circumstances of human placement with random variations. These annotations are then used to train a downstream automatic segmentation model in SAM-L.}
\label{fig:box}
\end{figure*}

\subsubsection{Tight bounding boxes}
We utilized the SAM model to streamline the annotation process. SAM enables both experts and lay annotators to draw bounding boxes~\cite{sochor2018boxcars} around regions of interest, eliminating the need to manually outline entire structures (it's worth noting that we also experimented with point annotations, but they did not yield useful pixel-level masks through SAM). This bounding box annotation approach is recognized as a more time-efficient method for experts, freeing them from the task of detailed delineation of objects. SAM then generates pixel-level segmentation from these box prompts, serving as annotations for training deep learning models. In this study, we generated tight boxes based on the pixel annotations we had previously gathered, to assess their efficacy.

\subsubsection{Random bounding boxes}
While tight boxes are directly created as minimal-size containers to envelop the original pixel-level annotation, humans often find it challenging to place such ``perfect" bounding boxes in real-world scenarios. The random boxes, as illustrated in Fig.\ref{fig:box}, emulate more realistic conditions in which humans place bounding boxes, allowing for random variations. Specifically, we added random offsets to the minimum and maximum coordinates of the tight bounding boxes, simulating the variability inherent to human-created bounding box annotations. Fig.\ref{fig:box} contrasts the tight box with a random one. For consistency during training, we used only one random offset box per image.

\textbf{The rationale for using random bounding boxes:}

(1) Regularization and Generalization. Random bounding boxes offer a form of regularization during training. By introducing random offsets to bounding box coordinates, the model becomes less sensitive to exact localized features. This approach mimics the variations observed in human box annotations fed into the SAM model, potentially enhancing segmentation performance, especially in cases with occlusions or partial object visibility.

(2) Increased Coverage. Random bounding boxes can be either slightly larger or smaller than the tight ones, effectively broadening the region surrounding the object. This proves advantageous when objects aren't perfectly centered within the ground-truth bounding box or when the mask has minor inaccuracies or noise. A larger coverage enables the model to assimilate more context, which bolsters segmentation accuracy.

(3) Robustness to Variations. Random bounding boxes capture the inherent variability seen in real-world object appearances and forms, thus bolstering the model's adaptability to differences in object size, orientation, and shape, ensuring it performs optimally even in demanding scenarios.

\subsection{SAM-L deep cell segmentation}
In our study, we incorporate a recent corrective learning strategy~\cite{deng2023democratizing} intended to boost robustness during the training of a segmentation model, especially within the realm of molecular imagery. Fig.\ref{fig:framework} details the entire labeling and auto-quantification process. Challenges emanate from a dearth of molecular expertise and inconsistencies in staining quality, resulting in unreliable annotations by non-experts. Drawing inspiration from confidence learning~\cite{northcutt2021confident} and similarity attention techniques~\cite{li2021dual}, our method selects the top-k pixel feature embeddings exuding higher confidence, representing pivotal regions for the present cell type. We then determine the cosine similarity between these chosen embeddings and a random pixel to emphasize areas of consensus between the model and lay annotations. Integrating these confidence scores into the loss function adeptly tackles noisy labels, enhancing the model's segmentation prowess for molecular images. This molecular-oriented corrective learning (MOCL) strategy~\cite{deng2023democratizing} offers a competent and potent means to process lay annotations, achieving results on par with models trained using expert annotations.

\subsection{Evaluation metrics}
F1 score is used to evaluate the performance of a binary classification model, such as distinguishing between diseased and healthy tissues or identifying tumors in medical images. It provides a balanced assessment of a model's ability to correctly classify both positive and negative cases. In medical imaging, the Dice score is often used to evaluate the accuracy of an image segmentation algorithm by comparing the algorithm's segmentation results with the ground truth manual segmentation. Higher Dice scores indicate better segmentation accuracy. It is a popular metric because it is sensitive to both false positives and false negatives, providing a balanced measure of segmentation performance.

\section{Data and Experiments}
\subsection{Data}
We obtained PAS images of glomerulus slides and annotation images, annotated by one experienced pathologist and three computer science students. In total, 11 WSIs underwent PAS staining, which included 3 slides featuring injured glomeruli. These WSIs were paired with the corresponding IF images to aid in subsequent procedures. The stained tissue samples were digitally scanned at a magnification of 20$\times$. After a thorough multi-modality multi-scale registration process, we successfully compiled and annotated a set of 1,147 patches containing podocyte cells, in addition to 789 patches with mesangial cells. Each patch measures 512$\times$512 pixels.

The dataset consists of patches derived from WSIs, with each patch depicting specific glomeruli or molecular structures. The dataset was randomly divided into three subsets: training, validation, and testing, in a ratio of 6:1:3, respectively. This partitioning ensured a balanced distribution of both injured and normal glomeruli across the subsets. Automated tuft segmentation and molecular knowledge images aided in the identification of glomeruli and cells. The annotation was consistently performed using ImageJ (version v1.53t). The ``Synchronize Windows" feature was employed to enhance the annotation process, allowing for a synchronized cursor display across modalities with spatial correlations. The ``ROI Manager" managed the resulting annotations, storing the binary masks for each cell type. The input data for SAM includes anatomical images paired with manual annotations provided by both experts and students. This diverse annotation strategy guarantees extensive coverage and enables the assessment of SAM's efficiency and accuracy in segmenting glomerular structures.

\begin{table}[t]
\caption{Mean annotation accuracy of three lay annotators with PAS+IF (F1 score)}
\centering
\begin{tabular}{lllllll}
\hline
\multirow{2}{*}{Method}                                                                                 & \multicolumn{2}{l}{Injured glomeruli} & \multicolumn{2}{l}{Normal glomeruli} & \multicolumn{2}{l}{Average} \\ \cline{2-7} 
 & Podocyte          & Mesangial         & Podocyte         & Mesangial         & Podocyte     & Mesangial    \\ \hline
\begin{tabular}[c]{@{}l@{}}Manual Contour  \\ (pixel-level annotation)\end{tabular} & 0.8374            & 0.8434            & 0.8619           & \textbf{0.8511}            & 0.8496       & \textbf{0.8473}       \\ \hline
\begin{tabular}[c]{@{}l@{}}SAM-L \\ (tight box annotation)\end{tabular}         & 0.8334            & 0.8279            & 0.8401           & 0.8335            & 0.8370      & 0.8312       \\ \hline
\begin{tabular}[c]{@{}l@{}}SAM-L \\ (random box annotation)\end{tabular}        & \textbf{0.8564}            & \textbf{0.8470}            & \textbf{0.8853}           & 0.8466            & \textbf{0.8577}       & 0.8469       \\ \hline
\label{table:mean}
\end{tabular}
\end{table}

\subsection{Experiments}
The experiment involved one experienced pathologist and three lay annotators who lacked specialized knowledge. We extracted glomerular anatomical and molecular patches from WSIs using a workstation equipped with a 12-core Intel Xeon W-2265 Processor and an NVIDIA RTXA6000 GPU. The cell contours were delineated on a separate workstation that featured an 8-core AMD Ryzen 7 5800X Processor and an XP-PEN Artist 15.6 Pro Wacom tablet. Annotating a single cell type on one WSI took approximately 9 hours, while staining and scanning 24 IF WSIs in a batch required about 3 hours. Both experts and lay annotators utilized an identical experimental setup, ensuring a fair comparison. This setup allowed us to evaluate the performance differences between the two groups, highlighting the potential advantages of using non-expert annotators to reduce labor and time without compromising the quality of annotations.

\section{Results}
\begin{figure*}[t]
\begin{center}
\includegraphics[width=1\linewidth]{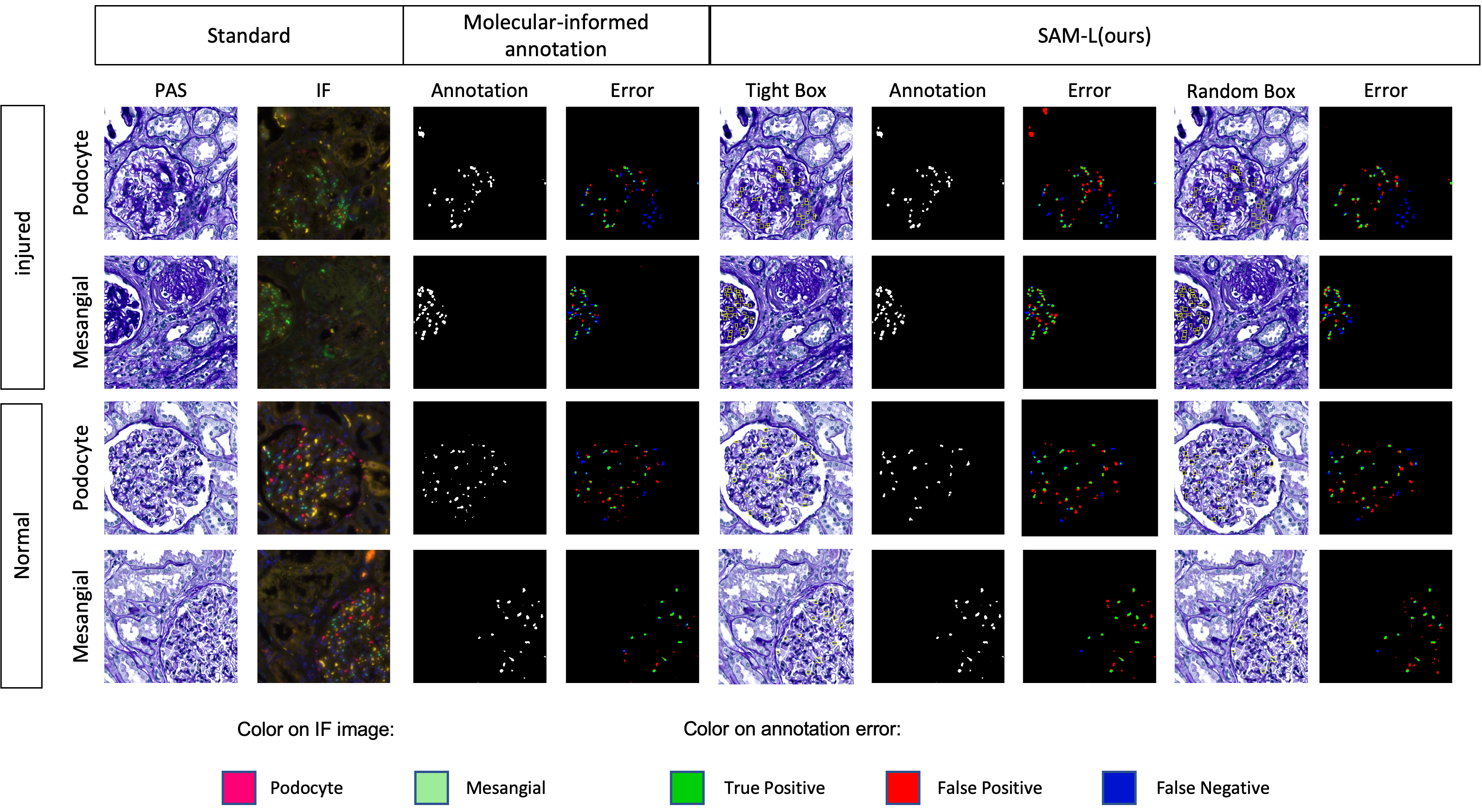}
\end{center}
\caption{Annotation accuracy using different methods. Note that the results were all based on annotations from an experienced renal pathologist.}
\label{fig:accuracy}
\end{figure*}

Fig.\ref{fig:accuracy} and Table.\ref{table:mean} display the annotation accuracy comparison between different methods used for segmenting both injured and normal glomeruli. The focus is specifically on the podocyte and mesangial cell types. The average F1 scores from two expert and three student lay annotators are presented.

\subsection{Performance on Downstream Multi-class Cell Segmentation}
In this section, we present the results of our proposed method, which combines SAM\cite{deng2023segment} and MOCL\cite{deng2023democratizing}. We compare these results with those of other approaches that use different types of annotations. Table.\ref{table:performance} summarizes the performance of deep learning-based multi-class cell segmentation for both injured and normal glomeruli, with a specific focus on the podocyte and mesangial cell types.


When utilizing SAM with tight bounding box annotations in conjunction with MOCL, we noted performance variations between student annotations and expert annotations. The F1 scores for podocytes (injured glomeruli) ranged from 0.7043 to 0.7105, and for mesangial cells (injured glomeruli) they ranged from 0.7014 to 0.7027. For podocytes (normal glomeruli), F1 scores ranged between 0.7390 and 0.7657, and for mesangial cells (normal glomeruli), they were between 0.6513 and 0.6683. The highest average F1 scores were 0.7362 for glomeruli and 0.6891 for mesangial.


To assess the impact of random bounding box annotations, we conducted experiments using SAM with random bounding boxes and MOCL. The F1 scores ranged from 0.7170 to 0.7226 for podocyte (injured glomeruli) and from 0.6994 to 0.7096 for mesangial (injured glomeruli). For podocyte (normal glomeruli), F1 scores ranged from 0.7565 to 0.7673, and for mesangial (normal glomeruli), they ranged from 0.6713 to 0.6723. The highest average F1 scores were 0.7434 for glomeruli and 0.6949 for glomeruli. 


\begin{table}[t]
\centering
\caption{Performance of downstream deep learning based multi-class cell segmentation (F1 score).}
\begin{tabular}{lllllllll}
\hline
\multirow{2}{*}{Methods}                                                       & \multicolumn{2}{l}{\multirow{2}{*}{Datasets}} & \multicolumn{2}{l}{Injured glomeruli} & \multicolumn{2}{l}{Normal glomeruli} & \multicolumn{2}{l}{Average} \\ \cline{4-9} 
& \multicolumn{2}{l}{}                          & Podocyte          & Mesangial         & Podocyte         & Mesangial         & Podocyte     & Mesangial    
\\ \hline

\multirow{2}{*}{\begin{tabular}[c]{@{}l@{}}MOCL \\ (pixel-level annotation)  \end{tabular}}   
& Experts    & \multicolumn{2}{l}{0.7124}   
& 0.7022	&0.7547	&0.6674	&0.7321	&0.6685       
 \\ \cline{2-9}                                                      & Students   & \multicolumn{2}{l}{0.7198}   
 & \textbf{0.7157 }    & 0.7657      & \textbf{0.6830}      & 0.7411    & \textbf{0.7028}       
\\
\hline

\multirow{2}{*}{\begin{tabular}[c]{@{}l@{}}SAM-L\\ (tight box annotation)\end{tabular}}   
& Experts  & \multicolumn{2}{l}{0.7105}   & 0.7027            & 0.7657           & 0.6683            & 0.7362       & 0.6891 
\\ \cline{2-9}        
& Students    & \multicolumn{2}{l}{0.7043}   
&0.7014	&0.7390	&0.6513	&0.7205	&0.6817           
\\ 
\hline

\multirow{2}{*}{\begin{tabular}[c]{@{}l@{}} SAM-L \\ (random box annotation)\end{tabular}} 
& Experts     & \multicolumn{2}{l}{\textbf{0.7226}}   & 0.6994            & \textbf{0.7673}           & 0.6713            & \textbf{0.7434}       & 0.6883 
\\ \cline{2-9} 
& Students     & \multicolumn{2}{l}{0.7170}   
& 0.7096	&0.7565	&0.6723	&0.7354	&0.6949   
\\ \hline

\label{table:performance}
\end{tabular}
\end{table}

\section{Conclusion}

In this paper, we evaluate the potential of bypassing pixel-level delineation by directly leveraging the recent SAM model on weak box annotations in a zero-shot learning manner. Our proposed SAM-L model utilizes the zero-shot SAM model to transform weak box annotations into robust pixel-level annotations, which in turn bolsters the accuracy and efficiency of cell segmentation in digital pathology. Experimental results indicate that SAM-L, when employing random box annotations, delivers performance comparable to traditional benchmarks that use pixel-level annotations. These findings highlight the SAM-L model as a promising tool that not only boosts accuracy but also expedites the annotation process. This can further catalyze the broader adoption and practical applicability of molecular-empowered learning in real-world scenarios. In essence, this study paves the way for streamlining and democratizing the annotation process for training pathological image segmentation models, allowing us to depend primarily on lay annotators.

\section{ACKNOWLEDGMENTS}       
This work has not been submitted for publication or presentation elsewhere. This work is supported in part by NIH R01DK135597(Huo), DoD HT9425-23-1-0003(HCY), and NIH NIDDK DK56942(ABF).

\bibliography{main} 
\bibliographystyle{spiebib} 

\end{document}